\documentstyle[honnef,pslatex,epsfig,equats]{article}  
\frompage{000} \topage{000}                                          
\def\rightmark{Baryon-Rich Quark-Gluon Plasma}
\def\leftmark{M. Danos et al.}
\title{Baryon-Rich Quark-Gluon Plasma in Nuclear Collisions$^a$}
\authors{
{Michael Danos$^{1,b,c}$ and Johann Rafelski$^{2,d}$ %
}\\[2.812mm]
{\normalsize
\hspace*{-8pt}$^1$National Bureau of Standards\\
Washington, D.C. 20234\\[0.2ex] 
\hspace*{-8pt}$^2$ Institute of Theoretical Physics and Astrophysics\\
Department of Physics, University of Cape Town\\
Rondebosch 7700, Cape, South Africa
}}
 
\abstract{The maximum achievable temperature (energy density)
 and minimum kinetic energy required for the formation of 
a baryon-rich quark-gluon plasma formed at central rapidity 
in small impact parameter nuclear collisions is estimated. 
A possible mechanism leading to the pile-up of matter is introduced. 
Plasma formation is expected to appear at about 15 GeV/Nucleon uranium 
beam energy on a stationary target or 2.7 GeV/Nucleon in colliding beams.}

\newcommand{\simleq}
{\mbox{\raisebox{-1.2ex}{$\textstyle \sim$}
 \raisebox{.05ex}{$\textstyle  \!\!\!\!\!\! <$  }}}

\newcommand{\simgeq}
{\mbox{\raisebox{-1.2ex}{$\textstyle \sim$}
 \raisebox{.05ex}{$\textstyle  \!\!\!\!\!\! >$  }}}

\newcommand{\simrho}
{\mbox{\raisebox{-.25ex}{$\textstyle \rho$}
 \raisebox{.75ex}{$\textstyle  \!\!\!\!\!\! \sim$  }}}

\renewcommand{\theequation}{\thesection.\arabic{equation}}

\begin{document}

\maketitle
\vspace*{24pt}

\section{Introduction}
Two extreme pictures of a high energy collision between two heavy nuclei 
suggests themselves:
\begin{itemize}
\item[(a)] collision between two rather transparent bodies where 
the reaction products remain essentially in the projectile and the 
target reference frames respectively,
\item[(b)] collision between two rather absorbent bodies in 
which matter piles up in the collision and where the reaction 
products appear in the central rapidity region. This is the 
scenario we study in this report.
\end{itemize}
Off hand picture (a) would seem to be the more reasonable one considering 
the rather small high-energy hadron-hadron cross sections. This is the basis 
of a number of models purporting to describe the high-energy nuclear collisions. 
However, recent experimental evidence from p-nucleus collisions and cosmic ray 
data indicate that case (b) is the more frequent channel for the reactions 
leading to quark-gluon plasma formation [1]: these 100 GeV p-nucleus experiments 
indicate according to the analysis of Busza and Goldhaber [2] that the pp-data 
seriously underestimate the extent to which heavy nuclei would slow each 
other down. They find that in traversing the other nucleus, a heavy nucleus 
would lose perhaps 2.8 units, instead of only one unit, of rapidity. 
Thus there would be nothing left of the central baryon-free region 
suggested by hadronic cascade calculations [3]. While this substantial 
collective slowing effect is experimentally established at 100 GeV 
laboratory energy collisions, the cosmic ray data indicate a similar 
phenomenon at ultra-high energies [4]. In particular, lower limits 
on cross sections associated with the high-entropy-producing very high 
particle multiplicity events can be deduced from these data. It is found 
that events in which also a substantial thermalization of the longitudinal 
motion must have occurred are seen with a frequency of at least 1\% of 
all hard ultra-high energy collisions. It can be convincingly argued 
that the observed high multiplicities
are most easily understood if the collision has passed through a 
stage consisting of the quark-gluon plasma since as we will see 
below the equations of state of the plasma phase indicate a 
substantial intrinsic entropy.

In order to explain the above Fermilab and cosmic-ray data, 
we propose here a mechanism which allows to us reconcile the 
nuclear transparency with the opacity needed for the generation 
of the plasma. This is accomplished by considering the consequences 
of a small high-density region, henceforth called  ``plasma seed''. 
Under certain conditions, to be further discussed here, such a plasma 
seed once formed can begin to grow by capture of the trailing nucleons
 of the colliding nuclei. This then could lead to a baryon-rich plasma 
in the CM-frame, i.e., in the central rapidity region.

The nature of the plasma seed will be described in some detail 
in section 4;  we assume it to be a momentary large fluctuation 
in particle (and energy) density over a volume of the order of 
the size of a nucleon. This, of course, implied that the bags 
of the participating nucleons have merged, and the quarks thus 
occupy effectively a single common bag. In short, we assume the 
seed to consist of a state similar in nature to the quark-gluon 
plasma, albeit small in size, with sufficiently thermalized momentum 
distributions and with some color de-confinement. However, chemical 
equilibrium between different particle species, 
i.e., quark flavors, is not required.

In sections 2 and 3 we estimate the rate of energy accumulation 
in the plasma. Our approach is based on the observation that in 
order for the plasma seed to lead to an extended plasma the 
energy supplied to it by the incoming nucleons from the projectile 
and target nuclei must be larger than that lost by the plasma seed. 
This, and in detail the energy gain, is discussed in Section 2.

As concerns the energy loss, particularly within the initial very 
short plasma formation time (of perhaps 5 $\times$ 10$^{-24}$ sec = 1.5 fm/c), 
it occurs, in our opinion, mainly through pion emission [5] from the surface 
of the seed. Namely, in our approach the growth of the plasma region 
is caused primarily by the (microscopic) creation of suitable conditions 
for the phase transformation and not by collective flow. Hence during the 
formation period we do not need to consider the cooling arising from surface 
motion. We describe this in detail in section 3.

In section 4 we discuss the characteristics a local fluctuation 
of particle density over the hadronic volume must have in order 
to be able to seed a plasma event. We further estimate the 
frequency of occurrence of such a fluctuation and show that 
the formation of the seed is not only an occasional, but 
indeed a relatively frequent event.

In section 5 a scenario for plasma evolution is established. 
We find that the transitory occurrence of the plasma state can 
happen already at the quite moderate energies of the order of 
3 GeV for each nucleon in the CM system of the nuclear collision. 
We also obtain kinematic and geometric constraints for the 
formation of a baryon-rich plasma in the central rapidity region. 
In this section we also derive the maximum temperature achievable 
in the most favorable case as a function of the available kinetic 
energy. As can be easily argued (c.f., section 2) the largest plasma 
will be formed when the seed arises early in the reaction, thus 
predominantly in the central rapidity region for symmetric 
collisions ($A_{p} = A_{t}$). (Note, however, that even in these 
events a non-negligible distribution towards projectile and 
target rapidity regions must occur.) In such events, the 
{\bf baryon number} content of the plasma will be appreciable 
for very large nuclei, {\bf with strong presence in the central 
rapidity region.} A different scenario arises in those events 
in which owing to the absence of the seed fluctuation, or the 
smallness of the projectile, the baryon-rich plasma state is 
not formed early in the collision. There the baryons would 
probably be found mostly in the projectile -- target rapidity 
regions, owing to the known substantial transparency of 
normal nuclei to high-energy particles [3]. Still, the high 
radiation energy density reached in such collisions could 
lead to a baryonless plasma in the central rapidity region. 
Our reaction channel must be viewed as a complementary but 
orthogonal mechanism as compared to these high transparency 
reactions.

Finally in section 6 we turn to the discussion of the plasma 
equations of state and the associated temporal plasma evolution, 
and describe in qualitative terms possible observables of plasma events.

\section{Central Plasma Formation}
While the plasma receives energy and baryon number by the 
nucleons impacting on it, it also inevitably loses energy. 
Thus in order to grow there must hold for the total plasma energy $E$,   
\begin{equation}
\frac{dE}{dt} \; = \; \frac{dE^{A}}{dt} \; - \; \frac{dE^{R}}{dt } \; > \; 0  
\end{equation}
where $dE^{A}/dt$ is the heating by the incoming nucleons absorbed 
in the seed, and $dE^{R}/dt$ is the energy loss (which we assume 
later to be dominant by thermal pion radiation). If $dE/dt$ is 
negative the plasma (seed) will fizzle rather than grow. We now 
discuss these two terms, beginning with the gain term.

The energy influx into the plasma is controlled by the nuclear 
four-velocity, $u^{\nu} = \gamma(1, \vec{v})$; the plasma 
surface normal vector as seen from the CM-frame, 
$n^{\mu} = (0, \vec{n})$; the nuclear energy-momentum 
tensor, $T_{\mu \nu}$; and the probability for the 
absorption of an incoming nucleon by the plasma, $a$. 
Thus we have, with $d^{2}A$ the surface element,
\begin{equation}
\frac{dE^{A}}{dt} \; = 
\; \displaystyle{\int} \; d^{2}A \left( - T_{\mu \nu} \; 
       u^{\mu} n^{\nu} a \right) \; \; .
\end{equation}
As is well known
\begin{equation}
T_{\mu \nu} \; = \; \varepsilon_{0} \; u_{\mu} \; u_{\nu} + p  \; 
      \left( u_{\mu} \; u_{\nu} - g_{\mu \nu} \right)
\end{equation}
where $\varepsilon_{0}$ and $\rho$ are the energy density and the 
pressure in the rest frame of the projectile or target nucleus, 
respectively ($p$ is included here for completeness only). Hence we have
\begin{equation}
T_{\mu \nu} \;  u^{\mu} \; n^{\nu} \; 
   = \; \rho_{0} \; m \; \gamma \; \vec{n} \cdot \vec{v}
\end{equation}
where $\rho_{0}$ is the equilibrium nuclear density, 
i.e., $\rho_{0} = 1/6 \;\mbox{fm}^{-3}$. We do not consider 
the influence of the likely increase of the energy and 
particle density of the projectile or target in their 
rest frames arising from the entrance channel interactions. 
In order to err on the conservative side we compute as if 
all of the interacting region would instantly turn into the 
plasma state without compressions of nuclear degrees of freedom; 
a possible increase of the densities would make the environment 
even more suitable for the occurrence of a plasma.

Returning now to the evaluation of the energy gain, 
we have in the CM frame in terms of the projectile 
laboratory energy per nucleon, $E_{p}$,
\begin{subequations}
\begin{equation}
v \; = \; \left( \frac{E_{p} - m}{E_{p} + m} \right)^{1/2}
\end{equation}
\begin{equation}
\gamma \; = \; \frac{(2E_{p} m + 2m^{2})^{1/2}}{2m} \; \; .
\end{equation}
\end{subequations}

The probability of absorption coefficient $a$ is assumed, as usual, to be
\begin{equation}
a(z) \; = \; 1 \; - \; e^{-z/\lambda}
\end{equation}
where $z$ is the thickness of the plasma region and 
$\lambda$ is the average absorption length of a hadron 
in the plasma. When weighted with $\vec{n} \cdot \vec{v}$ 
over the plasma surface this leads to
\begin{equation}
\bar{a}(R) \; = \; \frac{1}{2} \; \bigl\{ 1 + 2e^{-2R/\lambda} \; 
  \left[ \frac{\lambda}{2R} + \left( \frac{\lambda}{2R} \right)^{2}\right] 
         - 2 \left( \frac{\lambda}{2R} \right)^{2} \bigr\} \; \; .
\end{equation}
The overall factor 1/2 in (2.7) reflects the ratio between 
the surface of a circle with radius $R$ and a half sphere, 
for $\lambda/R \rightarrow$ 0. We note that the absorption 
coefficient $\bar{a}(R)$ is indeed the average {\bf absorption probability}. 
Through $\lambda$ it depends on the particle density in the plasma, 
i.e., the temperature and baryon density. Still, the gain term 
in Eq.\,(2.1) depends mainly on the projectile energy. 
The final expression is, in detail,
\begin{eqnarray} \nonumber
\frac{d^{3}E^{A}}{d^{2}Adt} & = & 
\frac{1}{2} \; \rho_{0} \; 
\left( \frac{E_{p} - m}{E_{p} + m} \right)^{1/2} \; 
      \left( 2E_{p} \; m + 2m^{2} \right)^{1/2}\\[.2cm]
          & \times & \frac{1}{2} \; \left[ 1 + 2 e^{-2R/\lambda} \; 
     \left( \frac{\lambda}{2R} + \left( \frac{\lambda}{2R} \right)^{2} \right) 
        - 2 \left( \frac{\lambda}{2R} \right)^{2} \right] \; \; .
\end{eqnarray}

\section{Energy Loss}
\setcounter{equation}{0}
We now turn to the description of the energy loss term of 
Eq.\,(2.1). Two mechanisms for the energy loss from a plasma have been
considered: viz., expansion of the plasma [3b] 
and particle radiation \cite{5.}. At least in the beginning, 
i.e., at the time of decision between grow and fizzle, the 
expansion should play no role as the impacting nucleons provide 
an inertial confinement for the plasma. However, pion evaporation 
from the plasma is still possible, and the cooling associated 
with this process provides the energy loss of Eq.\,(2.1). 
Of course, some of the emitted pions will be returned to the 
plasma by the incoming nucleons. However, this return will be 
too late to have an impact on the question fizzle or grow:  
once the process has fizzled, i.e., the plasma seed has hadronized, 
the collision is back to the hadron cascade regime. On the other hand, 
if plasma growth has taken place the returning pions will of 
course return their evaporation energy to the plasma and contribute 
to the ultimate energy density of the plasma. Also, the influence 
of the expansion has to be reconsidered then.

We now develop a quantitative model of the pion radiation 
suitable for surface temperatures of 150 - 220 MeV and 
moderate baryon densities, such that the particle density 
is less than $\sim$ 10 particles/fm$^{3}$. Under these 
circumstances surface collisions involving more than one 
particle per fm$^{2}$ are rare. Hence we can limit ourselves 
to consider sequential one-particle events. The basic physical 
process consists of a colored particle impinging on the boundary 
between the quark-gluon plasma and the physical vacuum. 
For very high-energy particles with energy above 1 GeV it 
is possible to view this process as a color flux tube breaking 
mechanism [6]. However, at $T  \sim$ 180 MeV most of the 
impinging particles have a mean energy centered around
 500 MeV and the mechanism of color flux tube breaking 
must be substantially improved to account for the quark 
binding effects, viz. the low pion mass.

Only the normal component of the momentum controls 
the emission process. The dominant emission effect 
associated with pions originates in an effective 
coupling of quarks or gluons to the pion field at 
the boundary. Such an effective quark-pion coupling 
is well known from the chiral bag models [7] and is 
given by the effective Lagrangian
\begin{equation}
L_{q \pi} \; = \; \frac{i}{2f} \; \bar{q} \; \gamma_{s} \; \vec{\tau} 
     \cdot \vec{\varphi}_{\pi} \; q \; \Delta_{s} \; \; .
\end{equation}
Here $\Delta_{s}$ is the surface $\delta$-function; 
and $f$ = 93 MeV is the pion decay constant. Equation\,(3.1) 
describes the following processes:
\begin{itemize}
\item[(a)] a quark or antiquark hits the plasma boundary 
and emits a bremsstrahlung pion while being reflected back;
\item[(b)] a quark-antiquark pair hits the surface and converts into a pion.
\end{itemize}
As the pion emission by the plasma surface is a direct process 
the resulting pion intensity spectrum is non-thermal: 
the spectral form is determined by the thermal {\bf quark} 
spectra [8]. But most importantly, the pion surface radiance 
can substantially exceed the black body limit set by a hot pion gas. 
We also note that since Eq.\,(3.1) is a representation of the 
nonlinear chiral coupling it can lead to radiation in excess 
of the quark black body limit; the radiation can reach the 
upper physical limit.

A similar effective force describes pion emission by 
gluons. The fact that here two gluons are needed does not 
{\it a priori} make this process any less important since 
one of them will be the low energy recoil gluon from the 
boundary, and it is known that such gluons have a strong 
coupling strength. The effective Lagrangian for this process 
is simply
\begin{equation}
L_{GG \pi} \; = \; \frac{1}{f_{G}^{\ 2}} \;  \; 
   \mbox{Tr} \; (\vec{E} \cdot \vec{B}) \;\; \varphi_{0} \; \Delta_{s} \; \; .
\end{equation}
Here $\varphi_{0}$ is the $\pi_{0}$ field, the only pion 
component to which couple the isospin breaking gluons. 
The coupling strength $f_{G}$ is not known experimentally 
since the gluon structure of the stable hadrons has not yet 
been unraveled. The proposed Lagrangian (3.2) describes 
the following processes:
\begin{itemize}
\item[(a)] a gluon hits the plasma boundary and radiates a 
bremsstrahlung $\pi_{0}$ while being reflected back into the plasma;
\item[(b)] two gluons meet each other at the surface and become a $\pi_{0}$.
\end{itemize}
Both Eqs.\,(3.1) and (3.2) must be viewed as semi-phenomenological 
expressions of the underlying microscopic processes for which the 
structure of the bag boundary plays an important role. Therefore, 
for the purpose of a qualitative estimate of the pion radiance of 
the plasma it is better to consider a purely kinematic model. The 
basic assumption here is that in order for the surface collision 
to lead to pion emission the particle momentum normal to the surface 
must exceed a certain threshold. In particular, this momentum has 
to be larger than the normal momentum of the emitted pion. We take 
this threshold momentum to be of the order of 1/4 GeV/$c$ which 
also accounts for the kinematic constraints imposed by the 
bremsstrahlung of the pion. Our results are quite insensitive 
to the precise choice, as well as to the actual shape of the 
threshold function $\theta$ describing the probability of pion 
emission. Hence we will use:
\begin{equation}
\theta (p) \; = \; \left\{ 
\begin{array}{l}
1, \;\;\; p_{\perp} \;\; \simgeq \;\;  p_{M} \; \sim  \; 1/4  \; \mbox{GeV}\\[.3cm]
0, \;\;\; 0 \;\; < \;\; p_{\perp} \;\; \simleq \;\;  p_{M}  \;\; \;\; .
\end{array} \right.
\end{equation}
We note that the mean energy (momentum) of the massless or practically 
massless particles is about 2.5-3 $T \sim$ 450-550 MeV and that the 
particle densities peak at $\sim$ 2 $T$. Hence almost half of all plasma 
particles can participate in the radiation cooling. The constraint (3.3) 
further implies that there is no influence of quantum statistics on the 
final states. This seems to be erroneous and one is tempted to argue that 
upon radiation of a pion the quarks must find a place in the phase space, 
while in contrast the gluons may show stimulated radiation. 
Still, no overall error ensues since both these effects compensate 
each other almost exactly, since the Bose and Fermi degrees of 
freedom at $\mu/T \sim$ 1 are about equal in number, and since 
for our purposes both quarks and gluons are massless. This 
presupposes, as Eq.\,(3.3) also does, that both quarks and 
gluons are about equally efficient in radiating pions.

The energy per unit surface and unit time that leaves the 
quark-gluon plasma is now simply given by
\begin{equation}
\frac{d^{3}E}{d^{2}Adt} \; = \; \displaystyle{\int} \; 
\frac{d^{3}p}{(2 \pi)^{3}} \; \rho (p) f (E) \; E(p) \; 
   \theta(p) \; \frac{d^{3}V}{d^{2}Adt}
\end{equation}
where $\rho (p)$ describes the phase space density of coloured particles
\begin{equation}
\begin{array}{lcl}
\rho (p) & = & g_{q} \left\{ \left[ \mbox{exp} 
\left( (p - \mu_{q})/T \right) + 1 \right]^{-1} + 
\left[ \mbox{exp} \left( ( p + \mu_{q})/T \right) + 1 \right]^{-1} \right\}\\[.3cm]
             & + & g_{G} \left[ \mbox{exp} (p/T) - 1 \right]^{-1} \;\; .
\end{array}
\end{equation}
Here $g_{q}$ is the quark degeneracy: 
$g_{q} = 3_{c} \times 2_{s} \times 2_{f}$ = 12 and $g_{G}$ 
is the gluon degeneracy $g_{G} = 2_{s} \times 8_{c}$ = 16. 
Since the energy $E(p) = |\vec{p}|$ leaving the plasma region 
is not the total energy contained in the leading particle we have 
in (3.4) included the efficiency factor $f$. Since two quarks form 
the emitted pion and a third particle carries back the excess 
color a naive degrees-of-freedom counting based on equipartition 
leads to $f$ = 2/3. The differential in Eq.\,(3.4) is simply 
the normal velocity of particles impinging on the plasma surface
\begin{equation}
\frac{d^{3}V}{d^{2}Adt} = \frac{d^{2}Adz}{d^{2}Adt} 
= \frac{dz}{dt} = v_{\perp} = \frac{p_{\perp}}{E(p)} 
= \frac{p_{\perp}}{(p_{\perp}^{2} + p_{\parallel}^{2})^{1/2}} \; \; .
\end{equation}

In view of the qualitative nature of our model it is sufficient 
to expand in Eq.\,(3.5) the quantum distributions and to retain 
only the Boltzmann terms for the $q \cdot \bar{q}, G$ distributions,
\begin{subequations}
\begin{eqnarray}
\rho(p) \; &\approx &\; \bigl[ g_{q} \; \eta(3) \; 2\, \mbox{cosh} \; 
( \mu/T) + g_{G} \; \zeta (3) \bigr] \;
e^{- \sqrt{p_{\parallel}^{2} + p_{\perp}^{2}}/T} \\
& \equiv &\; \frac{8}{3} \; g \; 
e^{\sqrt{p_{\parallel}^{2} + p_{\perp}^{2}}/T} \; \; ,
\end{eqnarray}
\end{subequations}
where we have corrected the counting of the Bose and Fermi degrees 
of freedom by inducing the phase space integral weights 
$\eta$(3) $\approx$ 0.9 and $\zeta$(3) $\approx$ 1.2 in 
Eq.\,(3.7a). Finally, we must still account for the 
requirement that the color and spin degrees of freedom 
of the emitting particles, i.e., the quarks or the gluons, 
be coupled to the quantum number of the emitted pions. 
This introduces a factor which is 3/8 for both cases. 
As we have already included this factor in the definition 
of $g$ for later convenience, we must correct here by 
showing a factor 8/3 in Eq.\,(3.7a). Collecting all factors 
we see that the effective number of Boltzmann degrees of 
freedom of quarks and antiquarks at $\mu_{q} = T$ is 12.5 
while that of gluons is 7.5. At $\mu_{q}$ = 0 the number of 
quark degrees of freedom (22) is about that of gluons. 
Thus $g$ varies between 16 and 21 as function of $\mu_{q}$.

Combining Eqs.\,(3.4) and (3.6) with Eq.\,(3.7) we obtain 
the generalized Stefan-Boltzmann law:
\begin{eqnarray}\nonumber
\frac{d^{3}E}{d^{2}Adt} & = & \bar{f} g \; 
\displaystyle{\int_{P_{M}}^{\infty}} \; \frac{dp_{\perp}}{2 \pi} \;
p_{\perp} \; \displaystyle{\int_{0}^{\infty}} \; 
\frac{p_{\parallel} dp_{\parallel}}{(2 \pi)^{2}} \;
e^{\sqrt{p_{\parallel}^{2} + p_{\perp}^{2}}/T}\\[.2cm]
       & = & \bar{f} \; \frac{g}{2 \pi^{2}} \; T^{4} \; 3\, e^{- P_{M}/T} \;
\left[ \frac{1}{3} \left( \frac{P_{M}}{T} \right)^{2} + 
\left( \frac{P_{M}}{T} \right) + 1 \right] \; \; .
\end{eqnarray}

In figure 1 we show the cooling rate calculated from Eq.\,(5.15) 
as a function of the surface temperature $T$, choosing $\mu_{q}/T$ = 1. 
For $\mu_{q}$ = 0 the values are lower by about 20\%. Our current values 
for the radiance of the plasma is at about half the rate given by us earlier 
in Ref.[5] where the pion radiation by gluons and the coupling to the pion 
quantum numbers had not yet been considered. From figure 1 we further see 
that the precise value of $p_{M}$, or, said differently, the precise form 
of the threshold function $\theta$, Eq.\,(5.10), does not matter. However, 
we note here that our estimate may be uncertain by perhaps a factor 2 
considering the qualitative nature of our considerations.
\begin{figure}[htb]
\vspace*{-1.2cm}
\begin{center}
\epsfig{width=7.5cm,angle=-90,figure=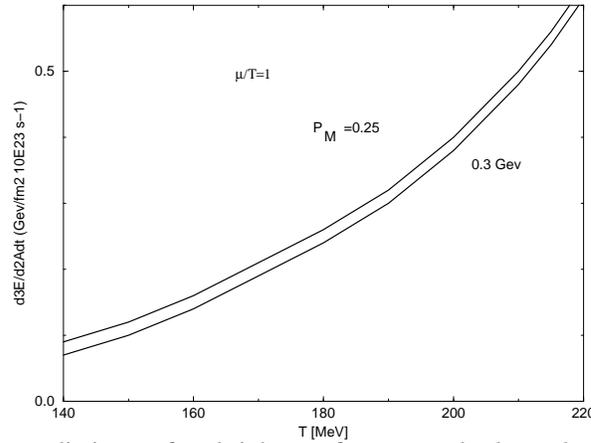}
\end{center}
\vspace*{-1.3cm}
\caption[]{
Pion radiation surface brightness from a quark-gluon plasma as
          function of temperature.
}
\vspace*{-0.5cm}
\label{fig1}
\end{figure}

We discuss briefly our results in terms of a numerical example 
chosen to represent a typical case of a quark-gluon plasma. 
Our example is a spherical plasma droplet of $R$ = 4 fm, a 
surface temperature of $T$ = 180 MeV, and $\mu_{q}/T$ = 1. 
The energy density then is 2.1 GeV/fm$^{3}$ according to the 
equation of state of a perturbative quark-gluon gas. Similarly, 
the baryon density is found to be $\sim$ 0.5/fm$^{3}$, i.e., 
about 3 $\rho_{0}$. The baryon number exceeds 150 if $T$ is 
somewhat larger in the interior than of the surface. Since 
0.7 GeV/fm$^{3}$ is needed for the creation of the final baryons 
implied by the assumed value of $\mu$, the available energy density 
is about 1.4 GeV/fm$^{3}$ and the total available energy is 
ca. 400 GeV. For this example we find for the rate of energy 
loss through the surface $A$,
\begin{eqnarray*}
\frac{dE^{R}}{dt} \; = \; A \; \bar{f} \; \mbox{0.25} \; 
\frac{\mbox{GeV}}{\mbox{fm}^{2}} \; \frac{c}{\mbox{fm}}
\; = \; A \; \mbox{0.5} \; \frac{\mbox{GeV}}{\mbox{fm}^{2}} \; 
\mbox{10}^{23} \; \mbox{s}^{-1} \; \; .
\end{eqnarray*}
We note that this confirms the assumption of a sequential 
individual-particle process: when one particle of 500 MeV 
impinges on a surface area of 1 fm$^{2}$ the next particle 
following it with light velocity would be behind by a 
distance of about 1 fm (i.e., several mean free paths). 
On the other hand, this indeed is a very large energy 
loss rate. In our example, the energy loss in the first 
10$^{-23}$ sec is ($A$ = 200 fm$^{2}$)
\begin{eqnarray*}
\Delta t \; \frac{dE}{dt} \; = \; \mbox{100 GeV} \; \; ,
\end{eqnarray*}
which represents a substantial fraction of the total 
available energy of about 400 GeV.

\section{The Seed}
\setcounter{equation}{0}
The basic phenomenon allowing for the existence of a seed is 
the fact that the particle density in the nuclear ground state 
is not uniform but has fluctuations, i.e., 
$\langle \rho^{2} \rangle - \langle \rho \rangle^{2} \neq$ 0. 
In fact, because of the spin and isospin of the nucleons up 
to four particles can occupy the same position. Actually, 
owing to the nuclear interactions it is unlikely that the 
corresponding maximal density will occur, but as we shall 
see, a density of two to three times $\langle \rho \rangle$ 
should be fully adequate for that fluctuation to act as a plasma seed.

There are two essential aspects in a qualitative description of 
the time development of the seed. The first is the probability 
that an incoming nucleon will find the seed; the second is the 
probability that a substantial fraction of the energy of the 
incoming nucleon gets trapped in the seed, i.e., that only a 
small fraction of the energy is scattered out from the 
interaction region. In order for this to happen we require 
that the quarks of the incoming projectile nucleon should 
undergo numerous interactions in the seed. It is quite likely, 
for example, that if during the collision with the seed each 
quark of the incoming nucleon undergoes about two or three 
scattering events with the quarks and gluons of the seed, 
most of the energy, and, in particular, the baryon number 
will have indeed been trapped. We will take this assumption 
as our starting point for the estimate of the stopping 
efficiency of the seed.

Notice here that any mechanism in which the nucleons 
do not get trapped in the seed is much less efficient 
in generating a plasma since the departing baryons take 
along a substantial fraction of the energy. Hence presumably 
the threshold for the production of a central rapidity 
baryon-less plasma will be much higher than that for our 
baryon-rich plasma.

Returning to our development, we begin by establishing 
the likelihood of a substantial density fluctuation. 
We will do this in terms of the non-interacting, i.e., 
uncorrelated gas model. Let $4 \pi/3 \; r_{s}^{3}$ 
be the neighbourhood volume which contains the centers of 
the nucleons making up the seed. Now recall the definition 
of the $N$-body density matrix for an $A$-body system:
\begin{equation}
\begin{array}{ll}
\simrho^{(N)} (x_{1} x_{2} \; \cdots  & x_{N}; \; y_{1} \; y_{2} \; 
\cdots \; y_{N}) \; = \; \displaystyle{\int} \; d^{3}x_{N+1} \; \cdots \; d^{3}x_{A}\\[.3cm]
              & \psi^{*} (y_{1} y_{2} \; \cdots \; y_{N} \; x_{N+1} \; \cdots \; x_{A}) \;
\psi(x_{1} \; \cdots \; x_{N}x_{N+1} \; \cdots \; x_{A})
\end{array}
\end{equation}
with the diagonal elements
\begin{eqnarray}
\rho^{(N)} \left( x_{1} \; \cdots \;  x_{N} \right) \; 
= \; \displaystyle{\int} \; d^{3}y_{1} \; \cdots \; d^{3}y_{N} \; 
\delta^{3} \left( x_{1} - y_{1} \right) \; \cdots \; 
\delta^{3} \left( x_{N} - y_{N} \right) \nonumber \\[.3cm]
\simrho^{(N)} \left( x_{1} \; \cdots \;  x_{N} \; ; \; y_{1} \; \cdots \; y_{N} \right) \; \; .
\end{eqnarray}
The density matrices are normalized to unity. We have for the number of 
$N$-body clusters present in the system within a volume $v$
\begin{equation}
W_{N} \; = \; 
\left( \begin{array}{ll} N\\
                                      A\\ \end{array} \right)    \; 
\displaystyle{\int} \; d^{3}x_{1} \; \cdots \; d^{3}s_{N} \; V_{v} \left( x_{1} \; 
\cdots \; x_{N} \right) \; \rho^{(N)} \left(x_{1} \; \cdots \; x_{N} \right) \;\; .
\end{equation}
Here
\begin{equation}
V_{v} \left( x_{1} \; \cdots \; x_{N} \right) \; = \; \left\{
\begin{array}{ll}
1 \; \; \mbox{if all } \; x_{i} \; \mbox{within the (seed) volume}\\[.3cm]
0 \;\; \mbox{otherwise}
\end{array}   \right. 
\end{equation}
and 
$\left( \begin{array}{ll} N\\
                                      A\\ \end{array} \right)$
is the total number of $N$-tuplets.

We now estimate $W_{N}$ in terms of the non-interacting gas model 
where the density matrix factorizes,
\begin{equation}
\rho^{(N)} \; \rightarrow \; \rho^{(1)} \left( x_{1} \right) \; \rho^{(1)} 
\left( x_{2} \right) \; \cdots \; \rho^{(1)} \left( x_{N} \right) \; \; ,
\end{equation}
and let us assume a uniform density ($r_{A}$ is the radius of the target nucleus)
\begin{equation}
\rho^{(1)} (x) \; = \; \left\{
\begin{array}{ll}
\frac{3}{4 \pi r_{A}^{3}} \; \; \mbox{inside the nucleus}\\[.3cm]
0 \; \; \mbox{otherwise} \; \; .
\end{array} \right.
\end{equation}
Then the probability for finding a given nucleon within a specified seed volume is
\begin{equation}
P \; = \; \left( \frac{r_{s}}{r_{A}} \right)^{3} \; \; .
\end{equation}
Herewith we can evaluate Eq.\,(4.3):
\begin{equation}
\displaystyle{\int} \; d^{3}x_{2} \; \cdots \; d^{3}x_{N} \; V 
\left(x_{1} \; \cdots \; x_{N} \right)
\; \rho^{(N)} \; \left(x_{1} \; \cdots \; x_{N} \right) \; = \; p^{N-1} \;
\rho^{(1)} \left(x_{1} \right)
\end{equation}
and, performing the remaining integral $d^{3}x_{1}$ we obtain
\begin{equation}
W_{N} \; = \; \left(  
\begin{array}{ll}
N\\
A\\  \end{array} \right) \; p^{N-1} \; \; .
\end{equation}
where of course surface effects have been neglected. Now define
\begin{equation}
\xi \; = \; \left( \frac{r_{s}}{r_{0}} \right)^{3}
\end{equation}
where $r_{0}$ = 1.2 fm is the nuclear radius parameter which is 
defined by $r_{A} = r_{0} \; A^{1/3}$. This gives for large $A$ and small $N$
\begin{eqnarray}\nonumber
W_{N} & = & \frac{A(A-1)(A-2) \; \cdots \; 
        (A-N)}{A^{N}} \; \frac{1}{N!} \; \xi^{N-1}\\[.3cm]
             & \approx & \frac{A}{N!} \; \xi^{N-1} \; \; .
\end{eqnarray}
The quantity $\xi$ can be written as a function of the baryon density 
in the fluctuation, $\rho$. Taking as a lower limit $\rho$ to be given 
by the assumption that the 3$N$ quarks are uniformly distributed over 
a sphere with radius $R = r_{n} + r_{s}$, were $r_{n} \approx$ 1 fm 
is the radius of a nucleon, we have
\begin{subequations}
\begin{equation}
\rho \; = \; N \;\rho_{0} \; \frac{1}{ \left[ \xi^{1/3} + (r_{n}/r_{0}) \right]^{3}}
\end{equation}
\begin{equation}
\xi \; = \; \left[  \left(  \frac{N\; \rho_{0}} {\rho} \right)^{1/3} 
\; - \; \frac{1}{1.2} \right]^{3}
\end{equation}
\end{subequations}
where $\rho_{0}$ is the normal nuclear density. 
Hence $W_{N}/A$ from Eq.\,(4.11) is an explicit function 
of the baryon density and is shown in figure 2. 
We note that for $A \approx$ 200 there is a probability 
of more than 10$^{-2}$ in each of the colliding nuclei 
to find a nine-quark cluster ($N$ = 3) at substantial 
compression. Once formed, the probability for being hit by 
one of the nucleons of the incoming other nucleus in a 
head-on nuclear collision is essentially unity.
\begin{figure}[htb]
\vspace*{-1.1cm}
\begin{center}
\epsfig{width=7.5cm,angle=-90,figure=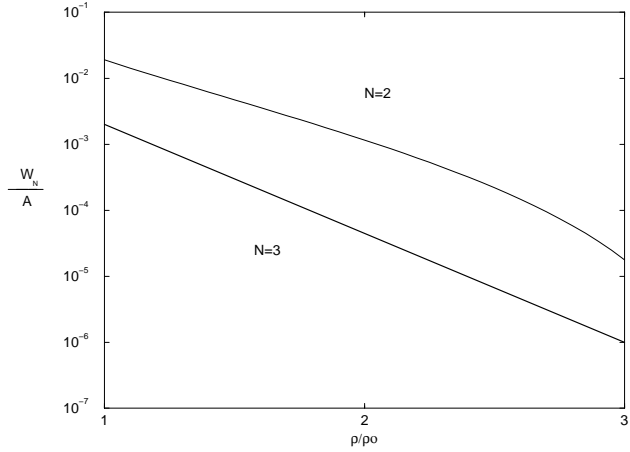}
\end{center}
\vspace*{-1.1cm}
\caption[]{
Probability per nucleon of a local density fluctuation for two-body
          ($N$ = 2) and three-body ($N$ = 3) clusters in the nuclear ground state.
}
\vspace*{-0.5cm}
\label{fig2}
\end{figure}

Having established quantitatively a lower limit for the likelihood of 
density fluctuations we now can consider the interaction length of an 
impinging nucleon. To that end we must discuss the role of hidden color 
in such a collision. Namely, in a collision of free nucleons the volume 
of the available final state phase space is reduced by the requirement 
that the reaction products be colorless. Without this requirement the 
quark-quark cross section would be about three times larger than the 
simple additive model estimate of 1/3 of the $p-p$ cross section, 
i.e. 10 mb. Since the color restrictions are relaxed in the seed, 
in particular in a three-nucleon seed, the quarks of the incoming 
nucleon there will have about a three times larger cross section. 
For the mean free path of a quark in the seed,
\begin{equation}
\ell \; \sim \; \frac{1}{\sigma_{q \; \mbox{\footnotesize{seed}}} \; 
                       \rho_{\mbox{\footnotesize{seed}}}} \; \; ,
 \end{equation}
taking therefore 
$\sigma_{q \; \mbox{\footnotesize{seed}}} \; \approx 
\; \mbox{3} \; \sigma_{qq} \sim$ 30 mb 
and $\rho_{\mbox{\footnotesize{seed}}} \approx$ two to three times 3$\rho_{0}$, 
we find $\ell \sim$ 0.4 fm. Recall that 3$\rho_{0}$ is the 
normal quark density in a nucleus.
An incoming quark will scatter on the average 2R/$\ell$ times 
in the seed of radius $R$,  i.e. five to six times. Consequently, 
the stopping distance $\lambda_{q}$ which we have associated with
 a total of five quark scattering lengths, i.e., about 2 fm is of 
the order of 2$R$ for the seeds described above. Note that as soon 
as energy has been deposited in the seed the particle density will 
increase substantially since gluons and $q \bar{q}$ pairs will be 
produced copiously. Thus the interaction length of relevance for our 
further considerations, i.e., once plasma has developed, is that of 
colored particles. Taking a conservative value of 15 mb for the 
average QCD cross-section, we find for a plasma at 2 GeV/fm$^{3}$ 
energy density, i.e., about 6 particles/fm$^{3}$ (each particle 
has c.a. 3$T$ $\sim$ 500 MeV energy in the plasma) a mean free 
path $\lambda \; < \;$ 0.1 fm.

The scenario we envisage is thus an initial accidental 
pile-up of energy in a small seed, followed by subsequent 
growth of the plasma by continued absorption of impinging nucleons.

\section{Details of the Plasma Formation}
\setcounter{equation}{0}
We now return to the discussion of the plasma 
formation condition: we set $dE/dt$ = 0 in Eq.\,(2.1). 
Given the energy gain Eq.\,(2.8) and energy loss Eq.\,(3.8) 
we find, for a given beam kinetic energy, the minimum size 
a plasma seed must have in order for it to grow. This is shown 
in figure 3 for a selection of plasma ignition temperatures, 
$T_{I}$, computed taking $\mu_{q}/T$ = 2. In the initial 
stages of the nuclear collision this is the probable value 
of the statistical parameters (see also section 6). Note 
that we err on the conservative side by enhancing the 
radiation losses by that choice.
\begin{figure}[htb]
\vspace*{-1.1cm}
\begin{center}
\epsfig{width=7.5cm,figure=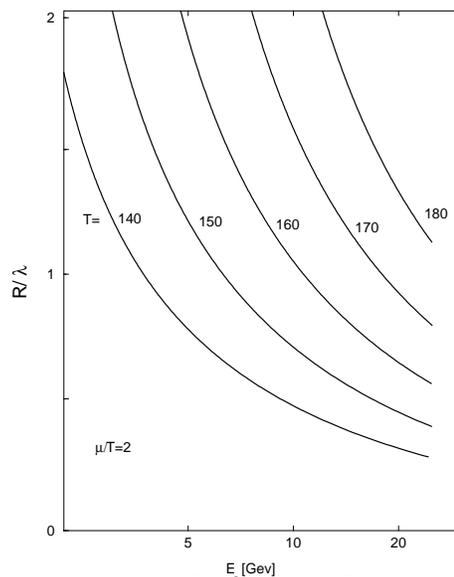}
\end{center}
\vspace*{-.1cm}
\caption[]{
Minimum size of a plasma seed as function of beam laboratory energy
            for different phase transition temperatures.
}
\vspace*{-0.5cm}
\label{fig3}
\end{figure}

For $R/\lambda$ $\sim$ 1 we notice that at 
$T_{I} \sim$ 150 - 160 MeV, a beam energy of 10 to 20 GeV/nucleon 
should suffice to lead to plasma ignition. (We have taken the 
seed size to be of the order of the nucleon size.) It seems 
rather unlikely that ignition can be achieved at much lower 
beam energies if the phase transition is of the first order. 
Thus below our beam energy limit a seed will very likely 
fizzle and we just achieve a superheated nucleon gas.

In order to understand the development of the collision process 
at sufficiently high beam energies after the initial plasma 
formation, one must re-examine the question of the cooling 
mechanism. First, there arises the possibility that the evaporation 
of pions may lead to a cooling off of the surface and hence to a 
shut-off of the evaporation process. Here the answer is found by
 considering the heat conductivity of the plasma which, if 
sufficiently large will maintain a surface temperature high 
enough for the pion radiation to continue. Second, it is possible 
that in later stages of the plasma other processes, in particular 
kinetic expansion, contribute to the cooling process. We discuss 
these two points in turn.

We begin by considering the heat conductivity. Since the 
plasma consists of rather free particles the naive expectation 
is that a sufficiently high conductivity will be available. 
Indeed, the basic relation between the heat flow $\vec{Q}$ 
and the energy $\varepsilon$ is
\begin{equation}
\vec{Q} \; = \; \ell \; \vec{\nabla} \; \varepsilon \left(T; \; \mu_{q}/T \right)
\end{equation}
where $\ell$ is the mean free path. Assuming that only a radial 
gradient of $T$ develops, with $\mu_{q}/T \sim$ const over the volume, 
the radiation equilibrium requires
\begin{equation}
\frac{d^{3}E^{R}}{d^{2}Adt} \; = \; Q_{r} \; 
= \; \ell \; \frac{\partial T}{\partial r} \; 
\frac{\partial \varepsilon}{\partial T} \; = \; \ell \; 
\frac{1}{T} \; \frac{\partial T}{\partial r} \; 4 \; \varepsilon
\end{equation}
where the last equality arises since $\varepsilon \; \sim T^{4}$. 
We now consider our numerical example. Taking the radiation loss as 
estimated in section 3, i.e., 
$\frac{1}{c} \; \frac{d^{3}E}{d^{2}Adt}$ = 0.17 GeV/fm$^{3}$ 
and the associated energy density as 2.1 GeV/fm$^{3}$, and 
assuming for $\ell$ a value in the range 1/3 - 1/5 fm we 
find that the required temperature gradient at the surface is,
$$
\frac{\partial T}{\partial r} \; = \; \frac{T}{\ell} \; 
\frac{{\mbox{0.17 GeV/fm}}^{3}}{4 \times {\mbox{2.1 GeV/fm}}^{3}} \; 
= \; (10 - 15) \; \frac{\mbox{MeV}}{\mbox{fm}} \; \; .
$$
It appears that this temperature gradient is just within sensible 
bounds, leading for a plasma radius of 4 fm to a temperature differential 
between the center and the surface of $\sim$ 50 MeV. We further note that 
unlike in nonrelativistic gases, the mean free path $\ell$ here is inversely 
proportional to $\partial \varepsilon/\partial T$ since it is inversely 
proportional to the particle density. Thus the above estimate can be made 
more precise. For $\mu_{q}/T \;<$ 2 the energy per particle in the plasma 
is just 3$T$ and hence the particle density $\rho = \varepsilon$/3$T$. 
Therefore we have
\begin{subequations}
\begin{equation}
\ell \; \frac{\partial \varepsilon}{\partial T} \; 
\approx \; \ell 4 \; \varepsilon/T \; 
=  \; \ell \; 12 \pi \; = \; 12/\sigma \; \; .
\end{equation}
Inserting this into Eq.\,(5.2) we obtain for the necessary temperature gradient,
\begin{equation}
\frac{\partial T}{\partial r} \; 
= \; \frac{d^{3}E^{R}}{d^{2}Adt} \; \frac{1}{12} \; \bar{\sigma}
\end{equation}
\end{subequations}
where $\bar{\sigma}$ is the average particle-particle cross section. 
The range of values given above for $\frac{\partial T}{\partial r}$ 
corresponds to 
$\bar{\sigma} \; \sim$ (5-10)mb, i.e., $\frac{1}{2}$ to 1 fm$^{2}$.

Secondly, we turn to the discussion of the kinetic expansion 
of the plasma. To begin with one must recognize that in contrast 
to the above discussed pion radiation process the collective 
expansion requires an organized many-body flow, i.e., a flow in 
which a hydrodynamic velocity is superimposed over the random 
thermal motion of all the quarks and gluons. Therefore the 
relevant time constant is given by the speed of sound and 
thus is about three times larger than the radiation time constant. 
Furthermore, the expansion is driven by the excess of the internal 
pressure over that exerted on the surface by the physical vacuum 
(see below). Now, the effect of the internal pressure on the surface 
is reduced by the pion radiation. The point is that those particles 
which are responsible for the pion emission do not exert their full 
force on the surface, or said differently, that emitted pions 
exercise a recoil force on the surface due to 
{\it actio paret reactionem} balancing. We now 
demonstrate how this relieves a substantial 
fraction of the internal surface pressure resulting 
from the particles impinging on the plasma surface. 
Balancing the momenta at the surface we find that when 
pion emission is allowed to occur the momentum recoil of the surface is,
\begin{equation}
\Delta p \; = \; \left\{
\begin{array}{rcl}
           2p_{\perp} & : & p_{\perp} \; < \; p_{M}\\[.3cm]
2p_{\perp} (1 - f ) & : & p_{\perp} \; > \; p_{M} 
\end{array} \right.
\end{equation}
where $f$ is the fraction of the normal momentum carried away by 
the emitted pion. We now re-compute the effective pressure on the plasma surface:
\begin{eqnarray}\nonumber
\bar{P} \;  &=&  \; g \; \left[ \; \displaystyle{\int_{0}^{p_{M}}} \; 
\frac{dp_{\perp}}{2 \pi} \; 2p_{\perp} v_{\perp} \; 
\frac{p_{\parallel} dp_{\parallel}}{(2 \pi)^{2}} \; \rho (p) \right.\\[0.3cm]
              & +& \; (1 - f) \; \displaystyle{\int_{P_{M}}^{\infty}} \; 
\left.\frac{dp_{\perp}}{2 \pi} \; 2p_{\perp} v_{\perp} \; 
\displaystyle{\int_{0}^{\infty}} \; \frac{p_{\parallel} dp_{\parallel}}{(2 \pi)^{2}} \; 
\rho (p) \; \right]
\end{eqnarray}
where we have used Eq.\,(3.6). Also, $g$ is the effective number 
of degrees of freedom for the particles in the plasma as defined
 below Eq.\,(3.7). We notice that the effective quark pressure $\bar P_{q}$ 
is equal to the expected quark pressure $P_{q}$ =1/3 $\varepsilon_{q}$, 
reduced by the contribution of high normal momentum particles, 
weighted by the factor $f$:
\begin{equation}
\overline{P} \; = \; P - fg \; \displaystyle{\int_{p_{M}}^{\infty}} \; 
\frac{dp_{\perp}}{2 \pi} \; 2p_{\perp} v_{\perp} \; 
\displaystyle{\int_{0}^{\infty}} \; 
\frac{p_{\parallel} dp_{\parallel}}{(2 \pi)^{2}} \;  \rho (p) \; \; .
\end{equation}
The important point to realize is that the contributions of particles with 
$p_{\perp} > p_{M}$ to the particle pressure $\bar{P}$ are dominant. 
To see this we evaluate, in obvious notation,
\begin{eqnarray}
\begin{array}{lcl}
\displaystyle{\frac{P(p_{\perp} > P_{M})} {P} }
& \equiv & 
\frac{\displaystyle{\int_{P_{M}}^{\infty}} dp_{\perp} p_{\perp}^{2} 
\displaystyle{\int_{0}^{\infty}} 
\frac{ p_{\parallel} dp_{\parallel} }
{ \sqrt{ p_{\perp}^{2} + p_{\perp}^{2} }  }
 \; \rho(p) }
{ \displaystyle{\int_{0}^{\infty} } dp_{\perp} p_{\perp}^{2} 
 \displaystyle{ \int_{0}^{\infty} } 
\frac{ p_{\parallel} dp_{\parallel} }
{  \sqrt { p_{\perp}^{2}  +  p_{\parallel}^{2} } }
 \; \rho(p) } \;
=  \; \frac{ \displaystyle{ \int_{P_{M}}^{\infty}} dp_{\perp} p_{\perp}^{2} \; 
e^{-p_{\perp}/T} }
{ \displaystyle{\int_{0}^{\infty} }  dp_{\perp} p_{\perp}^{2} \; 
e^{-p_{\perp}/T} } \\[1.5cm]
  & \; = \;  &  e^{-P_{M}/T} \; \left[ \frac{1}{2} \; 
\left( \frac{P_{M}}{T} \right)^{2} \; + \;
 \left( \frac{P_{M}}{T} \right) \; + \;  1 \right] \; \; .
\end{array}
\end{eqnarray}
This is a monotonically falling function of $P_{M}/T$; for 
$P_{M}/T \sim$ 1 - 1.5 we find that the ratio Eq.\,(5.6) 
varies between .92 and .81. Hence, inserting Eq.\,(5.7) 
into Eq.\,(5.6) we find for $f  \sim$ 2/3
\begin{equation}
\bar{P} \; = \; P \left( 1 - f \; \frac{P(p_{\perp} > P_{M})}
{P} \right) \; \cong \; 0.3 - 0.4 \; P \;\; .
\end{equation}
As Eq.\,(5.8) shows only about one third of the internal pressure 
acts on the surface. Thus, in effect, the time constant relevant for 
the cooling process through expansions is extended by a factor of 
about two. This leads to the conclusion that the kinetic expansion 
contributes only about 10-20\% to the cooling of the plasma. This 
effect is most pronounced for a baryon-rich plasma.

The physical and important difference between the effect of 
cooling of the plasma by pion radiation compared to cooling 
by kinetic expansion resides in that the former leads to a 
reduction of the plasma temperature {\bf without} a significant 
increase of the plasma volume. This, of course, has important 
consequences on the dynamics of the plasma development, and, 
in particular, on the observable quantities. For example, cooling 
by radiation will convert the internal energy more efficiently into 
pions than the expansion mechanism. In an expansion the energy is 
converted into collective motion and is manifested in the form of 
additional kinetic energy of the produced particles. Hence in the 
radiation cooling the available entropy is used to create more new 
particles, i.e., pions, while in the adiabatic expansion it is 
essentially contained in the kinetic motion. Therefore we expect 
that the high $p_{\perp}$ pion spectra will distinguish these processes.

Having established the likely dominance of pion cooling over 
kinetic expansion, we next discuss the maximally obtainable 
plasma temperature, neglecting the effect of the cooling by 
expansion. As already remarked, once the plasma has begun 
to grow a fraction of the radiated pions will be swept along 
by the incoming nucleons and re-enter the plasma. This process 
introduces a dependence of the loss term on the beam characteristics 
such as the baryon number. Even though this turn-around of the 
pions does not change the conditions for the plasma formation 
it influences greatly the maximal achievable plasma energy density. 
Since the thermal radiation is isotropic the returned fraction, 
$\eta$, will be of the order $\eta \; \simleq$ 1/2. To obtain 
an estimate of this maximum plasma energy density one has to 
multiply the energy radiation term, Eq.\,(3.8), with (1 - $\eta$) 
and balance it with the unmodified gain term, Eq.\,(2.8). 
We recall that in the derivation of of Eq.\,(3.8) a non-degenerate 
plasma gas has been assumed, and $\mu_{q}/T$ is expected to be 
less than 2. As the collision process continues the temperature 
of the plasma will grow until the nuclear collision terminates or, 
in case of heavy nuclear collisions (uranium on uranium), until 
the temperature has risen to a level at which the pion radiation 
(still proceeding sideways) overwhelms the (frontal) energy influx. 
This maximum achievable temperature is shown in figure 4 for a few 
choices of the pion turn-around coefficient $\eta$, as a function 
of projectile beam energy. In view of the high plasma density here 
we have used $R/\lambda$ = 5 (i.e., $R \sim$ 4-5 fm) and $\mu/T$ = 1 
(see next section). As one can see, the maximal temperature achievable 
in the collision does not depend too sensitively on the choice of the 
parameters and reaches for 50 GeV (i.e.,  5 GeV on 5 GeV in colliding 
beam U--U collisions) a value around 230 MeV. Hence, once a plasma
 has ignited, one can expect that a full-fledged quark-gluon plasma 
event will take place, with the energy density reaching 4-5 GeV/fm$^{3}$. 
We note again, that underlying this scenario is the requirement that 
the collisions take place between two quite heavy nuclei such that 
assumed $R$ plasma $\simgeq$ 2 $R$ nuclear. Only U--U collisions 
(or similar)qualify for this requirement.
\begin{figure}[htb]
\vspace*{-1.1cm}
\begin{center}
\epsfig{width=7.5cm,angle=-90,figure=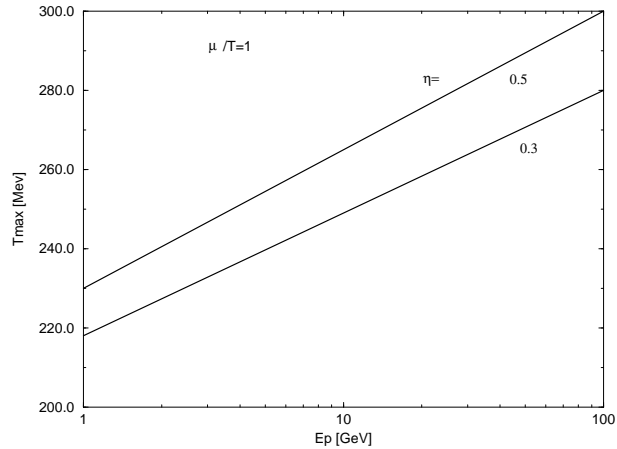}
\end{center}
\vspace*{-1.1cm}
\caption[]{
 Maximum achievable plasma temperature as function of beam laboratory
           energy for two values of the pion turn-around coefficient.
}
\vspace*{-0.5cm}
\label{fig4}
\end{figure}

After the end of the build-up phase, i.e., at the termination 
of the nuclear collision, the dynamics is governed by a complex 
process of pion radiation, hydrodynamic expansion, and surface 
hadronization of the plasma. At this point one must ask whether 
the density of the radiated pions is large enough for them to 
undergo multiple scattering, so that a pion gas cloud could 
be formed which would exert a back-pressure on the radiated 
pions, and thus could slow down the radiative energy loss 
of the plasma, and its hadronization.

We will not pursue this issue further here, 
except for an estimate of the number of pion-pion 
scattering processes. Consider the case where the
 emitted pions would form a density $\rho$ surrounding 
the plasma droplet of the form
\begin{equation}
\rho \; = \; \rho_{0} \left( \frac{R}{r} \right)^{2} \; \; .
\end{equation}
Let us consider that a given pion travels through a 
gas having the density distribution (5.9). In that case 
the scattering probability is given by ($j$ is the radial 
current density of the emitted pions which have survived up 
to the radial distance $r$ without having been scattered by 
the pion gas)
\begin{equation}
\frac{1}{r^{2}} \; \frac{d}{dr} \; (jr^{2}) \; 
= \; - j \sigma \rho \; 
= \; - j \sigma \rho_{0} \; \left( \frac{R}{r} \right)^{2} \; \; ,
\end{equation}
and hence we have
\begin{equation}
j \; = \; \frac{j_{0} R^{2} }{r^{2} } \; 
e^{- \sigma \rho_{0} R [1 - (R/r)]} \; \; .
\end{equation}
For $\sigma \rho_{0} R \ll$ 1 we find the unperturbed pion current 
which behaves like $r^{2}$. The exponential describes the effect of 
scattering in the gas. Taking for a numerical example 
$R \approx$ 4 fm, $\rho_{0} \approx$ 1 fm$^{-3}$, 
and assuming $\sigma \approx$ 0.2 - 0.5 fm$^{2}$, which is 
reasonable when recalling that the pion-pion scattering 
peaks at the $\rho$-meson mass, which is several line widths 
above a typical c.m. pion-pion energy, we find from the value 
of $j(r = \infty)$ which represents the unscattered part of the 
beam, that a pion will scatter one or two times on the way out to infinity.

\section{Plasma Properties}
\setcounter{equation}{0}
On several occasions we have had to assume the likely values 
of $\mu_{q}/T$ and $T$ in the plasma formed in nuclear collisions 
and it is obviously interesting to explore the path taken by an 
isolated quark-gluon plasma fireball in the $\mu_{q}-T$ plane, 
or equivalently in the $\nu-T$ plane. Several cases are depicted 
in figure 5. In the Big Bang expansion the cooling shown by the
dashed line occurs in a universe in which most of the energy is 
in the radiation. Hence, we have for the chemical potential: 
$\mu_{q} \ll T$. Similarly, the baryon density $\nu$ is 
quite small. In normal stellar collapse leading to cold 
neutron stars we follow the dash-dotted line parallel to 
the $\mu_{q}$-resp. $\nu$-axis. The gravitational compression 
is accompanied by (relatively) little heating.
\begin{figure}[htb]
\vspace*{-1.5cm}
\begin{center}
\epsfig{width=7.5cm,angle=-90,figure=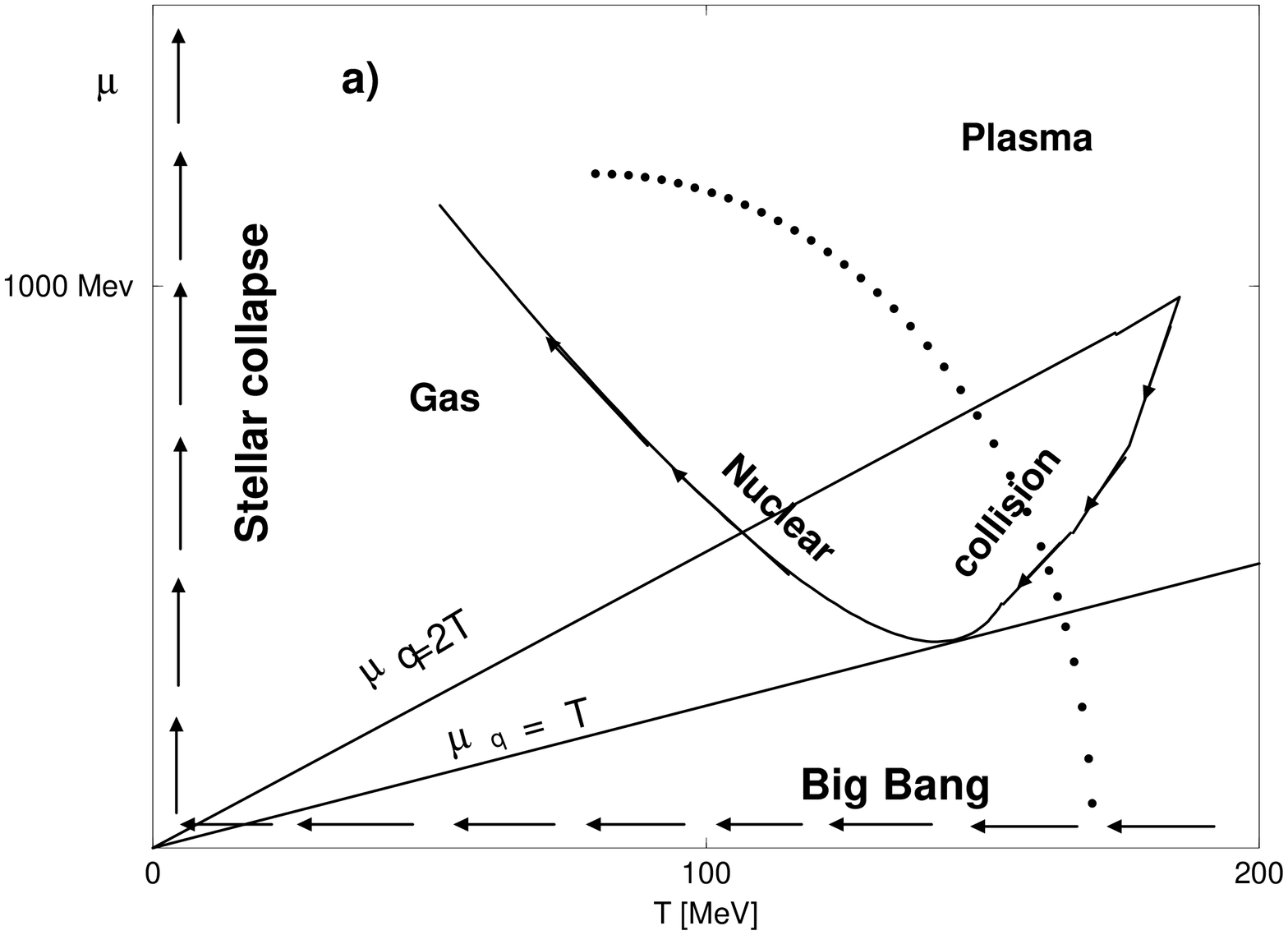}

\vspace*{-1.5cm}
\epsfig{width=7.5cm,angle=-90,figure=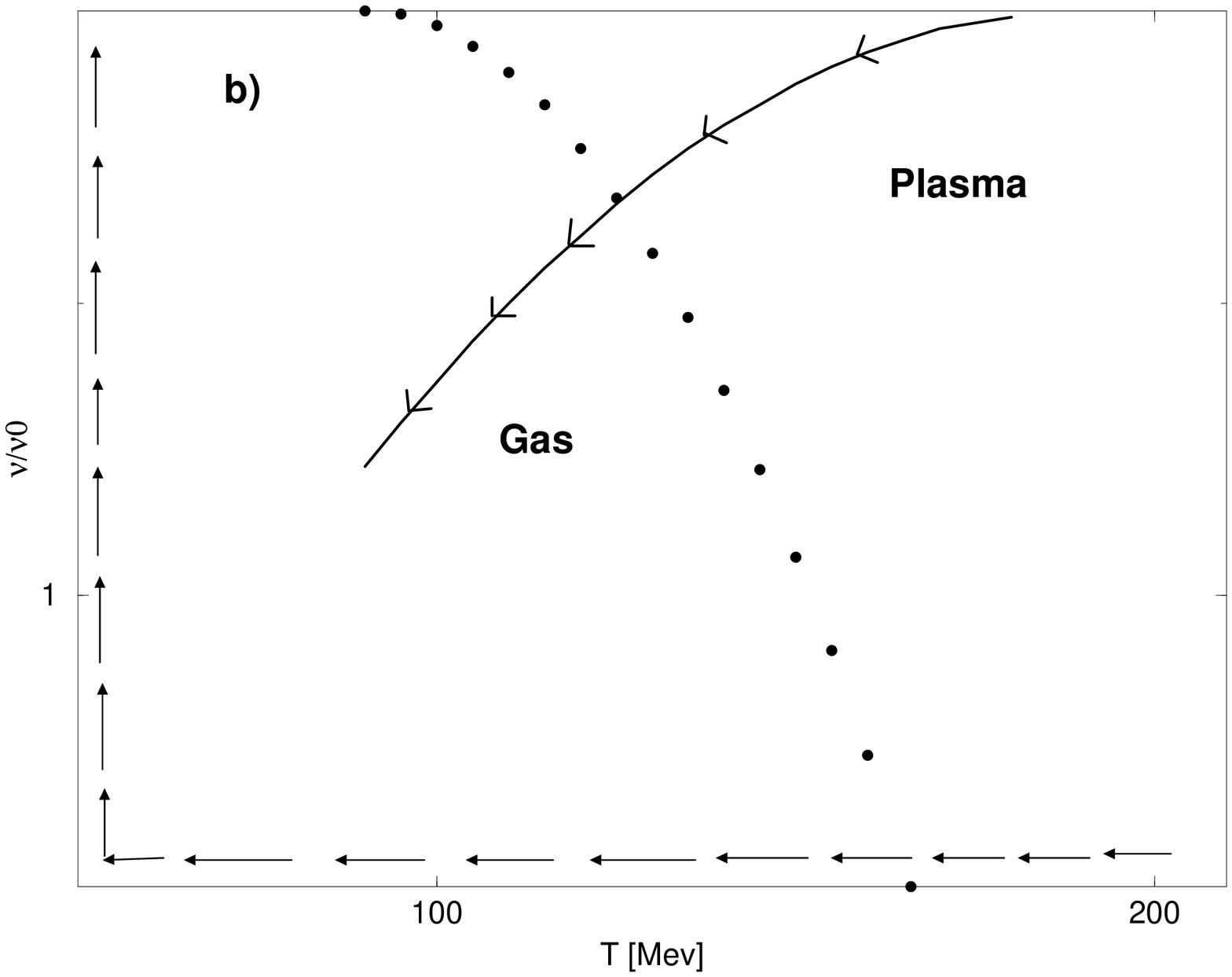}
\end{center}
\vspace*{-1.1cm}
\caption[]{
 Paths taken in the (a) $\mu_{q}-T$ plane and (b) $\nu-T$ plane by different
          physical events.
}
\vspace*{-0.5cm}
\label{fig5}
\end{figure}
In contrast, in nuclear collisions almost the entire $\mu_{q}-T$ (or $\nu-T$) 
plane can be explored by varying the parameters of the colliding nuclei. 
As we have already argued the most easily accessible region corresponds to
$\mu_{q}/T \; \simleq$ 2. To appreciate this further consider the baryon density
\begin{equation}
\nu \; = \; \frac{2 \pi}{3} \; \left( 1 - \frac{2 \alpha_{s}}{\pi} \right) \; 
(\delta^{2} + 1) \; T^{3} \delta \; \; ,
\end{equation}
where $\mu_{q}/ \pi T = \delta$. Since $\delta <$ 1 by assumption, 
we neglect $\delta^{2}$ against 1, that is
\begin{equation}
\nu \approx 1.4 T^{3} \; \delta \; \; , \; \; \frac{\mu_{q}}{\pi T} \; 
= \; \delta < 1  \; \; , \; \; \alpha_{s}  \; = \; .55 \; \; .
\end{equation}
At $T$ = 160 MeV we verify that (6.2) leads to $\nu$(160) = 3/4 $\delta$ fm$^{3}$. 
Hence $\nu$ = 2 $\nu_{0}$ implies $\delta$ = 2/9 in agreement with our 
prior assumption of a small $\delta$ ($\nu_{0}$ is the normal baryon 
density in nuclei, $\nu_{0}$ = 1/6 1/fm$^{3}$). Thus, as long as we are 
interested in the domain $T \simgeq$ 160 MeV, $\nu/\nu_{0} \; \simleq$
2.6 we are allowed to replace $\mu_{q}$ (i.e., $\delta$) by $\nu$ in the 
expressions for the entropy and energy density (Eqs.\,(4.34) and (4.37) 
respectively in Ref. [9]) by relating (6.2). We find explicitly:
\begin{equation}
s \; \approx \; 6.55 \; \frac{\nu^{2}}{T^{3}} \; + \; 8.12 \; T^{3}
\end{equation}
\begin{equation}
\varepsilon \; = \; 9.82 \; \frac{\nu^{2}}{T^{2}} \; + 
\; 6.1 \; T^{4} \; + \; B \; = \; \frac{3}{2} \; (6.55) \; 
\frac{\nu^{2}}{T^{2}} \; + \; \frac{3}{4} \; (8.12) \; T^{4} \; + \; B \; \; ,
\end{equation}
where in the last equality we emphasise the relation to Eq.\,(6.3), 
i.e., there is a factor $\frac{3}{4}$ in the radiation term, and a 
factor $\frac{3}{2}$ in the particle term between corresponding terms 
in entropy and energy density. It is interesting to note that at 
constant $\nu$ both $s$ and $\varepsilon$ have a minimum at the 
same value of $T$, which is
\begin{equation}
0 \; \; = \; \; \frac{ \partial s}{\partial T}{\big\vert}_{\nu} \; 
=  \;  \frac{\partial \varepsilon}{\partial T}{\big\vert}_{\nu} \; 
\rightarrow \;  T_{\nu} \; = \; 0.965 \; \; \nu^{1/3} \; 
= \;105 \; \mbox{MeV} \; (\nu/\nu_{0})^{1/3} \; \; .
\end{equation}
We recall that a plasma may be formed at $\nu \sim$ 2.5-3 $\nu_{0}$ 
in nuclear collisions leading to the minimum value
$T_{3 \nu_{0}}$ = 151 MeV, $T_{2.5 \nu_{0}}$ = 142 MeV. 
In the pion radiation evolutionary scenario of the quark gluon 
plasma we find thus find that both entropy density {\bf and} 
energy density decrease as the temperature decreases from its 
initial value around 200 MeV. This supports the proposition of 
pion radiation from the plasma at constant baryon density, until 
the minimum value $T_{\nu}$ of the temperature is approached.

We can further evaluate how much entropy each radiated pion removes 
from the plasma and how much must be generated in the radiation process. 
Here we recall that each pion carries away about 3$T$ MeV of energy and 
as a Boltzmann particle carries four units of entropy. From Eqs.\,(6.3) 
and  (6.4) we find that in a plasma 
\begin{eqnarray}
\begin{array}{lcl}
\frac s \varepsilon & \; = \; & \frac{1}{T} \; 
     \displaystyle{\frac{6.55 \; \frac{\nu^{2}}{T^{3}} \; + \; 8.12 \; T^{3}}
{ \frac{3}{2} \; 6.55 \; \frac{\nu^{2}}{T^{3}} \; 
+ \; \frac{3}{4} \; 8.12 \; T^{3} \; + \; \frac{B}{T}}   }\\[.75cm]
         & \; \approx \; &  \; \frac{4}{3T} \; 
 \left[ 1 \; -  \; \frac{6.55}{8.12} \; \frac{\nu^{2}}{T^{6}} \; + \; \cdots \;
    - 0.094 \; \frac{B}{T^{4}} \right]  \\[.5cm]
        & \; \approx \; & \; \frac{4}{3T} \; 
  \left[ 1 \; - \; 1.6 \; \sigma^{2} \; + \; \cdots 
\; - 0.094 \; (B^{1/4}/T)^{4} \right] \; \; .
\end{array}
\end{eqnarray}
Since $B^{1/4} \; \sim \; T$ and $\sigma^{2} \; \ll \;$ 1, 
to the precision of our approximation ($\sigma \; < \;$ 1) 
we find that when lowering the energy of the plasma by 3$T$ 
(i.e., by about the energy of the radiated pion) we lower 
its entropy content by about four units. As this is exactly 
the same as the entropy content of an emitted pion we conclude 
that the pion radiation is not a strongly entropy generating 
process, as it should be, in order for it to proceed without 
impediment.

The following additional useful information about the plasma 
can be extracted from Eq.\,(6.3). The total entropy is
\begin{equation}
S \; = \; sV \;=\; 6.55 \; \frac{b^{2}}{VT^{3}} \; + \; 8.12 \; VT^{3}
\end{equation}
where the baryon number $b = \nu V$. At fixed $b$ the 
minimum value of $S$ is at
\begin{subequations}
\begin{equation}
S_{\mbox{min}}{\bigg\vert}_{b=\mbox{const}} \; 
= \; 2 \sqrt{6.55 \times 8.22} \; b
\end{equation}
hence we find for entropy per baryon in the plasma
\begin{equation}
(S/b) \; > \; 14.6 \; \; .
\end{equation}
\end{subequations}
Since each pion carries away about 4 units of entropy, 
and since a nonrelativistic nucleon gas following the 
plasma event contains 
$(S/b)_{\mbox{gas}} \sim m_{N}/T  \; \simleq$ 7 units of 
entropy per baryon we find that a baryon-rich plasma event 
is characterized by a pion multiplicity which exceeds 
by at least a factor of two the number of participating baryons, i.e.,
\begin{eqnarray*}
n_{\pi}/b > 2 \; \; .
\end{eqnarray*}

Important information about the evolution of the plasma 
volume can be derived from the first law of thermodynamics 
($b$ is the baryon number)
\begin{eqnarray*}
dE \; = \; - PdV + TdS + \mu db \; \; .
\end{eqnarray*}
Since $db$ = 0 and $\Delta S$ = - 4 per emitted pion which 
carries the energy $\Delta E$ = - 3$T$, we find for the volume change
\begin{equation}
\Delta V \; = \; \frac{1}{P} \; (T \; \Delta S - \Delta E) \; 
= \; \frac{\Delta E}{P} \; \left( \frac{4}{3} - 1 \right) \; 
= \; \frac{1}{3} \; \frac{\Delta E}{P} \; \; .
\end{equation}
We see that given the reduction of the total energy by the energy 
of the emitted pion, the volume decreases by a well determined amount. 
Noting that
\begin{eqnarray*}
P \; = \; \frac{1}{3} \; (\varepsilon - 4B)
\end{eqnarray*}
we can write Eq.\,(6.9) as
\begin{equation}
\Delta V \; = \; \frac{1}{3} \; \Delta E /   \left( \frac{1}{3} \; 
\varepsilon \; - \; \frac{4}{3} \; B \right) \; 
= \; \frac{\Delta E}{\varepsilon - 4B} \; 
= \; - 3 \; \frac{T}{\varepsilon - 4B} \; \;  .
\end{equation}
For example, taking $\varepsilon - 4B = 3P = 1 $GeV/fm$^{3}$ 
and $T$ = 180 MeV we find that the emission of one pion 
($\Delta E \sim$ 500 MeV) reduces the volume by about 
0.7 fm$^{3}$. We believe that this small change in volume 
is compensated by the ongoing kinetic expansion of the 
plasma and have therefore taken $\nu$ to be essentially 
a constant in our qualitative considerations. 
Thus Eq.\,(6.2) describes at $\nu$ = constant the path 
taken in the $\mu_{q} - T$ plane in figure 5.

In view of the above discussion of the properties of the 
plasma phase which follow from the equations of state we 
believe that if the plasma continues to radiate energy in 
the form of pions from a roughly constant volume while the 
energy is being supplied from inside by heat conduction, 
this process may continue until such a time that the surface 
temperature drops below the transition point which is near 
to $T_{\nu}$ since there the entropy content of the plasma 
surface is lowest. More precisely, at 
$t$ = 1.5 $\times$ 10$^{-23}$ sec, we find (cf. section 3) 
that 150 GeV has been radiated by our plasma fireball of radius 
$R$ = 4 fm and the initial surface temperature of 180 MeV has 
decreased to $T$ = 150 MeV. This is close to the temperature 
of the transition to the hadronic phase and is a minimum of  
the surface entropy at $\nu$ = 3${\nu_{0}}$. Hence a possible, 
perhaps even likely, scenario is that in which the freezing-out 
and the expansion happen simultaneously. If that is the case, 
then the expanding hadronic gas may be quite dilute and a hadron 
will not undergo many scattering events before reaching an 
asymptotic distance, i.e., before becoming accessible to 
detection by an experiment. Of course, this would be ideal 
in that the detected particles then would reflect the properties 
and composition of the plasma at the freeze-out point 
(see also Ref. 10). These highly speculative remarks are 
obviously made in the absence of experimental guidance. 
A careful study of the hadronization process remains to 
be pursued, perhaps along these lines.

\section{Summary and Conclusions}
\setcounter{equation}{0}
The formation of a baryon-rich quark-gluon plasma 
appears to be an important reaction channel in collisions 
of heavy nuclei in the energy region of 2.5-5 GeV per nucleon 
in the center of mass frame of reference. In this paper we have 
explored the consequences of assuming a density fluctuation as 
a seed for the formation of plasma through the mechanism of 
energy pile-up arising from the trailing nucleons being 
absorbed in the initial density fluctuation.

In considering the evolution of the collision process 
we have distinguished between the initial formation 
phase (say the first 0.5 $\times$ 10$^{-23}$ sec) and 
the evolution phase (i.e., $\tau >$ 1.5 $\times$ 10$^{-23}$ sec). 
For the formation period, we have estimated the critical conditions 
for plasma formation by balancing the kinetic energy influx into 
the seed against the energy loss from the seed arising from 
particle emission into all directions. To estimate the plasma 
evolution we have then taken the energy loss by particle 
emission to be operational only sideways and assumed that 
the size of the colliding nuclei would be so large that 
matter influx would continue throughout the estimated 
life-time of the plasma (i.e., $R/c \; \geq \;$ 2 $\times$ 10$^{-23}$ sec). 
We thus focus on collisions of very heavy nuclei, such as 
uranium on uranium. Then we find that the maximum achievable 
temperatures exceed 200 MeV.

The subsequent plasma evolution is shown to be dominated 
by statistical pion radiation which establishes a temperature 
gradient of $\simleq$ 50 MeV between the center and surface 
of the plasma. At surface temperatures of about $T$ = 150 MeV 
the entropy density has a minimum at the fixed baryon density 
of $\nu \sim$ 3$\nu_{0} = \frac{1}{2}$ baryons/fm$^{3}$ and 
we expect hadronization of the plasma to ensue. We argue that 
since the statistical particle emission relieves much of the 
pressure between the surface of the plasma and the vacuum,
 no substantial kinetic flow has a chance to develop, and 
assume the volume of the plasma to stay roughly constant 
until the hadronization period.

The position we take is in disagreement with some recent 
investigations; the latter however neglect to consider the 
reduction of pressure on the plasma boundary arising from 
the pressure reduction associated with the emission of the pions. 
In particular we recall that at the surface
\begin{equation}
P \; = \; P_{q\footnotesize{G}} - B
\end{equation}
where $P_{q\footnotesize{G}}$ is the quark gluon pressure. Even if 
$P_{q\footnotesize{G}} \sim$ 4$B$ 
(i.e., $\varepsilon_{q\footnotesize{G}} \sim$ 12$B$), 
i.e., three times the equilibrium energy density of plasma, 
there would be scarcely enough pressure to balance the vacuum 
pressure once, as expected, 3/4 of $P_{q\footnotesize{G}}$ is relieved by the 
statistical pion emission. Thus, if the pion emission is at the 
level we estimate, there is no doubt that we should be more worried 
about  ``implosion'', rather than explosion of the baryon-rich plasma; 
here the word  ``implosion'' is used in the sense that the phase
 boundary would be moving inwards.

Quite aside from our model of the pion emission, which remains 
to be confirmed by more microscopic approaches, we wish to 
emphasize here again that the quark-gluon plasma state is an 
extremely entropy-rich state of matter which requires a large 
number of final state pions as carriers of this entropy, the 
minimum number being two per baryon emitted from the plasma 
(see also Ref.10). It would seem that the mechanism of pion
 radiation is just the valve necessary to reduce the entropy 
content of the plasma state in order to facilitate the 
transformation into the  ``entropy-poor'' hadronic gas phase -- 
into which the plasma finally must evolve.

In view of the presented calculations and discussions it 
appears almost certain that a 5 GeV Uranium Collider would 
permit the investigation of the properties of the baryon-rich 
quark-gluon plasma.

\section*{Notes}
\begin{notes}
\item[a]
This  manuscript has never been published. 
It was widely circulated  was  University of Cape Town
preprint UCT-TP 7/84 in November 1984. 
It comprises material of an earlier shorter manuscript by 
M. Danos and J. Rafelski, {\it Formation of quark-gluon plasma 
at central rapidity}, a Frankfurt University preprint 
UFTP-82/94, (December  1982), 10pp.  In order to preserve
the historic accuracy  only corrections of a few 
equation typos seen in UCT-TP 7/84, and 
update of footnotes and  references, were made.
\item[b] Deceased,  August 30, 1999, see:
http://physics.arizona.edu/\~{ }rafelski/MDOB.htm
\item[c] Visiting Scientist, 
Institute of Theoretical Physics and Astrophysics, 
University of Cape Town, 
Rondebosch 7700, Cape, South Africa.
\item[d] Permanent address since 1987: 
Department of Physics, University of Arizona, Tucson, AZ 85721,
E-mail: Rafelski@Physics.Arizona.EDU
\end{notes}

\vfill\eject
\newpage
\end{document}